\documentclass[12pt]{iopart}

\usepackage{iopams}  
\usepackage{graphicx}  
\begin{document}

\title{Polymeric forms of carbon in dense lithium carbide 
}

\author{Xing-Qiu Chen,$^{1,*}$ C. L. Fu,$^{1}$ C. Franchini$^2$}
\address{
$^1$ Oak Ridge National Laboratory, 
Materials Science and Technology Division, Oak Ridge, TN, 37831, USA \\
$^2$ Faculty of Physics, University of Vienna and 
Center for Computational Materials
Science,  A-1090 Vienna, Austria\\
$*$ Current address: Shenyang National Laboratory for Materials Sciences,
Institute for Metal Research, Chinese Academy of Sciences,
Shenyang, 110016, P. R. China
}
\ead{corresponding author: xingqiu.chen@imr.ac.cn}
\begin{abstract}
The immense interest in carbon nanomaterials continues to
stimulate intense research activities aimed to realize
carbon nanowires, since linear chains of carbon atoms
are expected to display novel and technologically relevant
optical, electrical and mechanical properties.
Although various allotropes of carbon (e.g., diamond,
nanotubes, graphene, etc.) are among the best known materials,
it remains challenging to stabilize carbon in the
one-dimensional form because of the difficulty to suitably
saturate the dangling bonds of carbon. Here, we show through
first-principles calculations that ordered polymeric carbon
chains can be stabilized in solid Li$_2$C$_2$ under moderate pressure.
This pressure-induced phase (above 5 GPa) consists of
parallel arrays of twofold zigzag carbon chains embedded in
lithium cages, which display a metallic character due to
the formation of partially occupied carbon
lone-pair states in \emph{sp}$^2$-like hybrids. 
It is found that this phase remains the most favorable one 
in a wide range of pressure. At extreme pressure (larger the 215 GPa)
a structural and electronic phase transition towards
an insulating single-bonded threefold-coordinated carbon network is predicted.
\end{abstract}

Atomic carbon is stabilized in innumerable molecular
configurations (allotropes) characterized by different
combinations of $sp^3$, $sp^2$ and $sp$ hybrid
orbitals\cite{castro}. The most well-known allotropes of carbon
are diamond with its $sp^3$ bonds, graphite, graphene,
fullerene and nanotubes with $sp^2$ hybrides, as well as
transitional forms of carbon with mixed hybridizations  such as
amorphous carbon. 
The properties of carbon nanomaterials are the subject of an intensive 
research activity, motivated by several reasons ranging from their
potentially revolutionary technological applications to more fundamental issues such as
the understanding of interstellar clouds and C-based superconductivity. 
In this context, linear carbon chains have gained a renewed attention
in recent years since they are considered precursors in the formation of carbon nanomaterials.
However, despite the intensive experimental effort,
the synthesis of ideal solid carbon constituted by linear
chains of carbon (polyynes or cumulenes) is still a major challenge\cite{bum,Sun,kim}.
According to the most recent literature, three different processes can lead to the
synthesis of atomic wires of carbon: (i) by removing carbon atoms from graphene using energetic electron 
irradiation\cite{jin09}, (ii) by assembling carbon cluster on graphitic nanofragments\cite{ravagnan} and
(iii) by chemical, electrochemical, and other sophisticated techniques as discussed in a recent review\cite{cataldo}.
Within a simplified view, all linear conjugated polymers with a chain-like structure may be
considered as one-dimensional (1D) carbon phases in which
dangling bonds of carbon are suitably saturated. This is the
case, for example, for polyacetylene $\rm (C_2H_2)_n$, where
each carbon is covalently bonded to one hydrogen and is
connected to the two neighboring carbon atoms by alternating
single and double bonds \cite{Shirakawa-2001}. The alternation
of bonds and the resulting insulating groundstate
is induced by a Peierls-like instability. Massive doping,
however, lifts the Peierls instability and leads to a highly
conducting metallic regime\cite{Chiang-1977, Roth-1995}.

Like hydrogen, lithium forms several binary compounds with
carbon,\cite{Sangster} but none of them reveal 1D features
similar to $\rm (C_2H_2)_n$. On the carbon-rich side the
best-known materials are lithium-graphite intercalation
compounds such as $\rm LiC_6$ \cite{Dahn}, whereas on the
Li-rich side the only compound which can be produced directly
from the elements is $\rm Li_2C_2$, characterized by triple
bonded C$\equiv$C dimers.\cite{Juza-1967, Ruschewitz-1999}
Considering that charge is donated from Li to C in Li$_2$C$_2$,
one might expect that Li$_2$C$_2$ under pressure could be
transformed to a polyacetylene-like structure with 1D
characteristics. This expectation is based on two reasons: (1)
the pressure decreases the lattice spacing and, therefore,
increases the interaction between C$\equiv$C dimers, and (2)
the role of the C-H bond in C$_2$H$_2$ could be replaced by
that of the lone-pair orbital in Li$_2$C$_2$ in the
stabilization of a chain-like structure.  

In this study, we applied density functional theory (DFT)
to perform an extensive search for possible
structural candidates for $\rm Li_2C_2$ under pressure.
Over more than 200 crystal structures were explored with
different local environment of carbon atoms. 
We find that by application of low pressure (5 GPa)
triple bonded C$\equiv$C dimers in Li$_2$C$_2$ can be transformed
into metallic carbon linear chains encapsulated in lithium cages.
The pressure-dependent evolution of this structure show that the carbon chains 
are stable over a wide range of compressions, and are eventually converted into 
insulating single-bonded C-C cubic networks above 215 GPa.
Thus, our computational experiment reveals that it is possible to 
construct conductive linear-chains of carbon by simply compressing solid Li$_2$C$_2$
at pressures accessible to modern high-pressure technology. 
Considering that Li$_2$C$_2$ can be easily produced and that, unlike polyacetylene, the conductivity
arises naturally upon pressure without the need of any doping treatment, we believe that 
our study will stimulate prompt experimental investigations in the foreseeable future.

First-principles calculations were mainly performed using the
Vienna \emph{ab initio} Simulation Package (VASP) \cite{G1}
with the ion-electron interaction described by the projector
augmented wave potential (PAW) \cite{G2}. We used the
generalized gradient approximation within the
Perdew-Burke-Ernzerhof (PBE) parameterization scheme \cite{G3}
for the exchange-correlation functional.
Brillouin-zone integrations were performed for
special k points according to Monkhorst and Pack technique. The
energy cutoff for the plane-wave expansion of eigenfunctions
was set to 500 eV. Optimization of structural parameters was
achieved by the minimization of forces and stress tensors.
Highly converged results were obtained adopting a very high
energy cutoff of 500 eV for the basis sets, and utilizing a
dense $12 \times 12 \times 12$ $\vec{k}$-point grid for the
Brillouin zone integration.
For the proposed Bmmb phase, phonon spectra under pressures have
been performed using density functional perturbation theory as implemented
in the Quantum-ESPRESSO code \cite{QE}, using norm-conserving pseudopotentials.

\begin{figure}
\centering
\includegraphics[clip,width=0.42\textwidth]{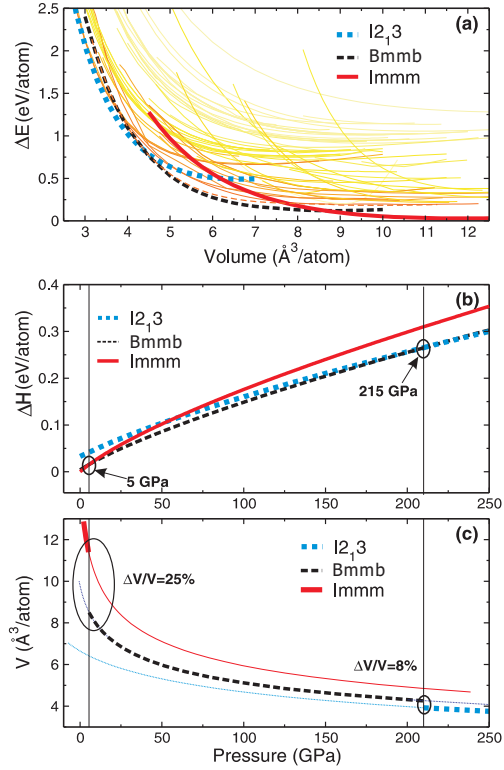}
\caption
{DFT calculated energy/pressure relations for Li$_2$C$_2$.
(a): energy difference ($\Delta$E) vs. volume ($V$),
(b): enthalpy difference ($\Delta$H) vs. pressure,
and (c): volume ($V$) vs. pressure.
In panels (a) and (b), the energy zero refers to the energy of the
\emph{Immm} phase at 0 GPa. The thin curves in panel
(a) represent the relationship
between $\Delta$$E$ and $V$ for other structures considered here. The thick
and thin curves in panel (c) denote the stable and unstable pressure
regions for corresponding phases, respectively.
}
\label{fig:s1}
\end{figure}

{\em Structural properties: }
the prediction of crystal structure
by means of first-principles methods remains
a complex task and a good choice of initial configurations
is crucial to perform an accurate structural search.
Recent studies have demonstrated
that a large pool of randomly selected initial configurations 
is a necessary ingredient for an accurate structural
search\cite{pickard,oganov06,oganov09}. In our study, to
form random unit cells we have chosen Bravais matrix with
random lattice parameters and random cell angles (including all
seven crystal classes). For each initial configuration the
internal degrees of freedom (atomic positions) were selected by
a random modification of several structural possibilities found
in literature.
Finally, the unit cell volume was properly scaled in order to reproduce the
desired pressure. At this point we have performed a full
structural optimization (cell shape, lattice parameters and
internal positions) for each starting configuration according
to a quasi-Newton algorithm where forces and the stress tensor 
are used to determine the search directions for
finding the equilibrium configuration.
This computational procedure reduces
substantially the risk of being trapped in metastable solution
(local minima of the potential energy surface) and permits the
exploration of a large number of different structures. We have
tested local carbon environments including triple bonded
dimers, carbon linear chains (polyyne- and
polyacetilene- like structures) and three-connected
nets\cite{Zheng} [planar (graphite like), puckered
($\alpha$-arsenic structure) or more complex conformations such
as those observed in $\rm ThSi_2$\cite{th}, $\rm
SrSi_2$\cite{sr} and $\rm DyGe_3$\cite{dy}.

\begin{figure}
\centering
\includegraphics[clip,width=0.42\textwidth]{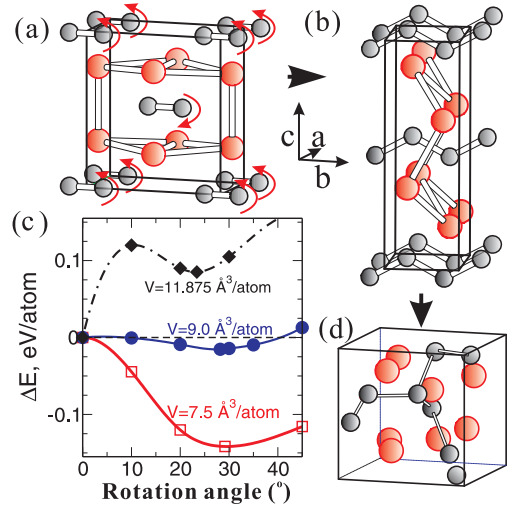}
\caption
{Schematic illustration of the pressure-induced structural
sequence (a) Immm  $\rightarrow$ (b) Bmmb  $\rightarrow$ (d) I2$_1$3
 with smaller balls for C and larger balls for Li.
Curved arrows in panel (a) indicate the rotation of  C$\equiv$C pairs which
leads to the zigzag C-C arrangement sketched 
in panel (b). Panel (d) illustrates the
threefold carbon polymeric network.
Panel (c) shows the change in energy ($\Delta$E)
when rotating the C$\equiv$C pairs as shown in
panel (a) for three different volumes of the  Immm structure.
Note that the lattice parameters are allowed to relax
(but with the constraint of constant volume)
in obtaining each of the curves in (c).
}
\label{fig:2}
\end{figure}

The energy/pressure relations derived from our DFT structural search
are depicted in Fig. \ref{fig:s1}.
The low-pressure insulating orthorhombic Immm phase 
[Fig. \ref{fig:2}(a)] is correctly predicted to be the lowest-energy
phase at zero pressure, with structural parameters in
excellent agreement with experiment (see
Table \ref{tab:s1}). 
At 5 GPa this phase is transformed into a
metallic phase with Bmmb symmetry characterized by
zigzag chains of carbon atoms which are embedded in
hexagonal Li cages [Fig. \ref{fig:2}(b)]. 
The Bmmb phase remains the most favourable solution until 215 GPa, 
at which pressure 
an insulating phase of cubic $\rm I2_13$ symmetry with the threefold
carbon polymeric network [Fig. \ref{fig:2}(c)] is stabilized.
The pressure-induced structural evolution is accompanied by a
peculiar modulation of the carbon bonds from C$\equiv$C triply
bonded dimers (Immm) to $sp^2$-like twofold-coordinated
carbon chains (Bmmb), and finally to a $sp^3$-like
threefold-coordinated carbon network, which is very similar to
that in the high-pressure cubic gauche structure of solid
nitrogen (cg-N) \cite{Chen}.

\begin{table}
\caption{DFT optimized structural parameters for the Immm
(No. 71), Bmmb (No. 63) and $\rm I2_13$ (No. 199) structures of
Li$_2$C$_2$. Lattice parameters (\emph{a}, \emph{b}, and
\emph{c}) and bond length of carbon ($L$$_{\texttt{C-C}}$) are
given in \AA.
\label{tab:s1}
}
\begin{tabular}{lrcccc}
\hline
Parameters   & Expt \cite{Ruschewitz-1999}&  DFT (0 GPa) &DFT (5GPa) & DFT (215 GPa) \\
\hline
\emph{a} &       3.652  &3.635 &3.134 &4.172   \\
\emph{b} &       4.831  &4.849 &2.503 &  \\
\emph{c} &       5.434  &5.389 &7.237 & \\
$L$$_{\texttt{C-C}}$ & 1.226 & 1.221 &1.431 &1.597 \\
Li &      4$j$:(0,0.5,0.236)& 4$j$:(0,0.5,0.239) & 4$c$:(0,0.25,0.153) & 8$a$:(0.306, 0.306, 0.306)\\
C &       4$g$:(0,0.127,0) &  4$g$:(0,0.126, 0) & 4$c$:(0,0.25,0.452) & 8$a$:(0.073, 0.073, 0.073)\\
\hline
\end{tabular}
\end{table}

\begin{figure}[h]
\centering
\includegraphics[clip,width=0.47\textwidth]{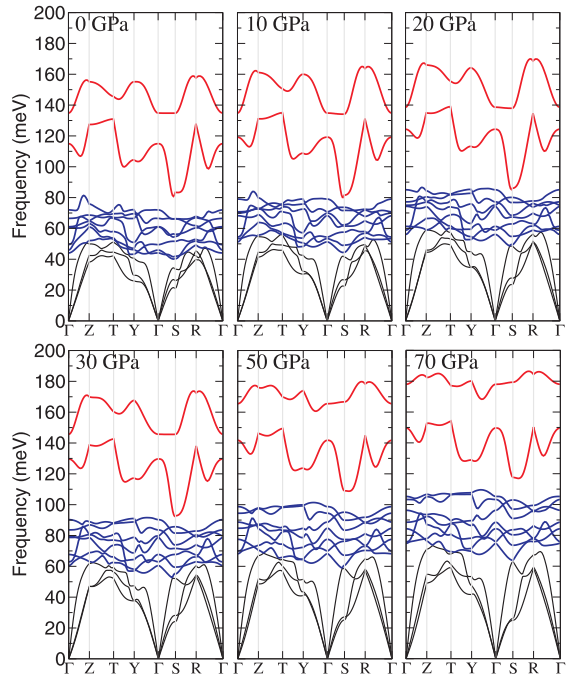}
\caption{Pressure-dependent phonon dispersions for the 
proposed Bmmb phase at the pressures of 
0, 10, 20, 30, 50, and 70 GPa.} 
\label{fig:s2}
\end{figure}

At 5 GPa, we find a sudden volume 
collapse of 25\%, as shown in Figure 
\ref{fig:s1}(c),
related to the Immm $\rightarrow$ Bmmb phase transformation.
The Bmmb phase is found to be energetically
stable against a phase separation into bcc Li and LiC$_6$, as
well as against a decomposition into bcc Li and graphite. 
Here, we should mention that the {\em ab initio} treatment of graphite 
within a conventional GGA approach suffers from the well-known limitations
related to the fact that van Der Waals (vdW) interactions between graphic layers 
are poorly described. In order to minimize the effects of vdW errors we have 
followed the procedure tested by N. Mounet and N. Marzari\cite{Marzari} which consists
in the calculation of the equation of state (EOS) in the variable c/$a_0$ with the 
in-plane lattice parameter $a_0$ kept fixed to the experimental value.
The so obtained EOS and bulk modulus are in fair agreement with the
experimental values\cite{expt1, expt2} and the expected error associated with 
the GGA binding energy, $\sim$ 0.05 eV/C, does not affect the alloy stability reported here.
Additional support for the stability of the proposed Bmmb structure is provided
by the absence on any imaginary frequency in the calculated pressure-dependent 
phonon spectrum reported in Fig. \ref{fig:s2}.

The predicted pressure-induced structural evolution from the
Immm phase to the Bmmb structure can be ascribed to the
modification of the lattice parameters (elongation of $c$
and compression of the $ab$ plane of the Immm phase) and
concomitant rotations of the carbon dimers, according to 
Fig. \ref{fig:2}(a) and (b). Energetically, the crucial
mechanism governing the Immm to Bmmb transition is the rotation
of the carbon dimers, as illustrated in Fig. \ref{fig:2} (c).
At zero pressure (corresponding to a volume of V=11.875
\AA\,$^3$) the rotation of dimers is highly unfavorable,  but
with increasing pressure the energy barrier is reduced and
disappears at a volume of V=9.0 \AA$^3$, corresponding to 5
GPa. Upon this structural transformation, the C$\equiv$C dimers
are converted into zigzag carbon chains with a C-C bond length
of 1.43 \AA\, and a large inter-chain distance of 3.13 \AA.
The carbon chains in the Bmmb phase are parallel
arrays of polymeric chains along the \emph{b} axis direction
[Fig. \ref{fig:2}(b)]. Although the Bmmb phase is calculated
to be the most stable in a very wide pressure range of 5 to 215
GPa, at pressures above 150 GPa it becomes very close in energy to
a tetragonal phase of $\rm I4_1/amd$ symmetry. 
In this $\rm I4_1/amd$ phase
the carbon atoms also form zigzag chains, 
but with blocks of chains oriented 90$^\circ$ with
respect to each other, similar to the Ge-network 
of BaGe$_2$ \cite{Evers}. 
At 215 GPa
we predict a second structural transition towards an insulating
cubic  $\rm I2_13$ phase [see Fig. \ref{fig:2}(d)], 
accompanied by a volume collapse of 8\%.
This phase has a $\sigma$-character
single-bonded C-C bond length of 1.60 \AA\ at 215 GPa, which
resembles that of diamond at 0 GPa (1.55 \AA). This phase
is similar to the high-pressure cubic gauche structure of solid
nitrogen (cg-N \cite{McMahan}) as experimentally synthesized by
Eremets {\em et. al}, \cite{Eremets} and as theoretically
studied by us\cite{Chen}.

{\em Electronic properties:} 
at zero pressure the Immm phase exhibits a wide gap
of 3.9 eV which slightly decreases upon pressure, reaching the
value of 3.7 eV at the critical pressure of 5 GPa, at which the
transition towards the metallic Bmmb structure
occurs. Upon further compression, the metallic regime is
finally destroyed at 215 GPa, at which pressure the most stable
$\rm I2_13$ phase shows a gap of 2.7 eV.
The calculated DFT gaps of insulators are
generally too small. By performing a screened exchange hybrid density
functional calculation for Li$_2$C$_2$ within the
Heyd-Scuseria-Ernzerhof scheme
we obtain a significantly larger gap (5.8 eV) for the Immm phase at 0 GPa.

\begin{figure}
\centering
\includegraphics[clip,width=0.49\textwidth]{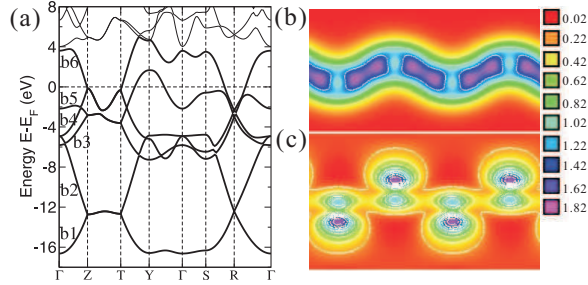}
\caption
{Panel (a): DFT band structures  of the Bmmb
phases; band labeling, see text.
Charge density contours of (b) C-C $\sigma$ bonded
states of bands b1, b2
and (c) lone-pair states of bands b3, b5.
}
\label{fig:4}
\end{figure}

\begin{figure}[h]
\centering
\includegraphics[clip,width=0.47\textwidth]{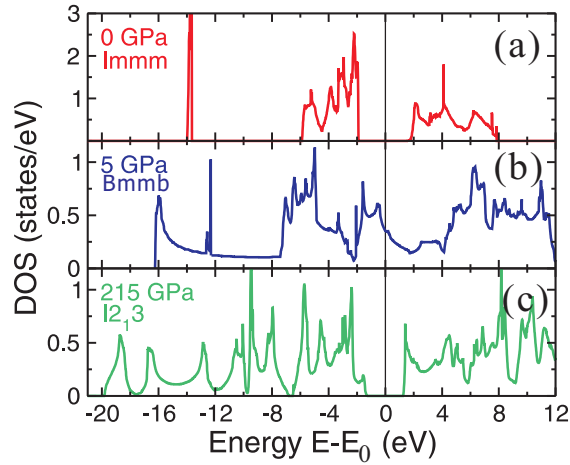}
\caption{Calculated density of states (DOS) for the Immm (0 GPa),
Bmmb (5 GPa), and $I2_13$ (215 GPa) structures.
For the Immm and I2$_1$3 phases, the energy zero is set in the middle of the gap;
For the Bmmb phase, the energy zero is set at the Fermi level.}
\label{fig:5}
\end{figure}

The bonding property of the metallic Bmmb structure is
dominated by \emph{sp}$^2$-like hybrids. Two of the carbon
electrons form covalent $\sigma$-type bonds [bands b1 and b2 in
Fig. \ref{fig:4}(a)] connecting neighboring carbons along the
zigzag backbone with an bond angle of 123$^\circ$. These
$\sigma$-character states [Fig. \ref{fig:4}(b)] display
quasi-1D features as reflected by a square root singularities
\cite{Roth-1995} in the density of states profile 
[Fig. \ref{fig:5}].
Our analysis also revealed
lone-pair orbitals localized at the carbon sites, as depicted
in Fig. \ref{fig:4}(c). The bands formed by the lone-pair
orbitals, however, are overlapped in energy with the bands
formed by the $\pi$-type bonding and antibonding orbitals of
carbon. In order to accommodate ten electrons per Li$_2$C$_2$
formula unit (with the charge donated from Li to C) in the
energy bands of Li$_2$C$_2$, the Fermi level has to intersect
both the lone-pair and $\pi$-state bands. In other words, there
are holes in the lone-pair state and electrons in the
antibonding $\pi$-state. The electron-deficient lone-pairs are
distributed into the fully occupied band b3 and partially
filled band b5. Formally, a lone-pair state should trap two
electrons but we find that about 1.7e are trapped in band 5.
The remaining electrons occupy the $\pi$-states, which
are characterized by the charge-density lobes aligned
perpendicular to the plane defined by the zigzag chains. These
orbitals form bands with $\pi$-type bonding and antibonding
character, as indicated by the respective bands b4 and b6 in
Fig. \ref{fig:4}(a). The bonding b4 band is fully occupied
whereas the antibonding b6 band is partially filled with
occupancy of 0.3e.

\begin{figure}
\centering
\includegraphics[clip,width=0.42\textwidth]{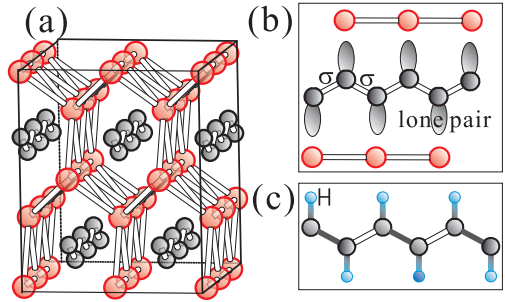}
\caption
{Panel (a): Supercell of the Bmmb
structure corresponding to panel (b) of
Fig. \ref{fig:2} with small balls for C and
large balls for Li.  Panel (b): view of
a $\sigma$ bonded C-C chain (bond length 1.45 \AA\,)
in which each C atom
carries a lone-pair
(the more delocalized $\pi$ bonds are not shown here).
Panel (c): similar view of polyacetylene C$_2$H$_2$
\cite{Shirakawa-2001}, in which each C atom is
connected by a double-bond
(length 1.36 \AA\,) and a single-bond
(length 1.44 \AA\,) to the neighboring
C-atoms. The C-H covalent bond length is 1.02 \AA\,.
}
\label{fig:6}
\end{figure}

As mentioned already, conjugate polymers represent the only
known producible form of carbon-chains based structures. In
order to highlight the structural similarities between
archetypal linear carbon chains of polyacetylene and C-C chains
of the Bmmb structure, we compare their corresponding local
structure in Fig. \ref{fig:6}. In spite of some visual
similarity, significant differences can be realized, which
explain the distinctly different electronic character: the
covalent nature of the C-H bond in polyacetylene is replaced by
the ionic C-Li bond in $\rm Li_2C_2$, which is stabilized by
the formation of the lone-pair in \emph{sp}$^2$-like hybrids.
In polyacetylene a Peierls distortion is observed
\cite{Shirakawa-2001} which induces alternating single and
double bonds with an angle of 120$^\circ$ and a gap opens up, 
and substatial doping is required in order to lift the system towards a conducting regime.
A different stabilization mechanism is found for the Bmmb phase
of $\rm Li_2C_2$: holes are created in the lone-pair band b5
which prevents the Peierls-like splitting of states, and
thereby the system is metallic.

To ascertain the distinct structural and electronic properties of the predicted
Bmmb phase, we have carried out additional calculations
with four different supercells containing a larger number of atoms
(Li$_{16}$C$_{16}$ - 2a$\times$2b$\times$c,
Li$_{24}$C$_{24}$ - 2a$\times$3b$\times$c,
Li$_{32}$C$_{32}$ - 2a$\times$2b$\times$2c, and
Li$_{64}$C$_{64}$ - 2a$\times$4b$\times$2c
) and performing the full structural relaxation 
without any symmetry constrains.
Indeed, all derived optimized structures converge to the same Bmmb phase
and display identical structural (carbon linear chains) and electronic (metallic 
\emph{sp}$^2$-like hybrids) characteristics.
This computational experiment unambiguously prove that the linear C-C chains
in the Bmmb phase are conducting and do not undergo any Peierls-like dimerization.
Also, it is interesting to note that there is a resemblance between
our proposed Li$_2$C$_2$ Bmmb structure
and Li$_x$B ($x$= 0.90 and 0.95) \cite{Cerrada}.
In the latter, each boron attracts one electron
from Li (thus becoming isoelectronic
to carbon), and B-B chains are formed,
encaged in a subnetwork of sixfold Li$_6$ octahedra.
However, in Li$_x$B the bonding picture is very similar to that
calculated for the hypothetical carbon chains
($\alpha$-carbyne or $\beta$-carbyne), which
is still intrinsically different from the current case
of Li$_2$C$_2$.

The metallic character of compressed Bmmb-type Li$_2$C$_2$ is maintained
until the occurrence of the second  pressure-induced transition
at 215 GPa, when the cubic $\rm I2_13$ phase is stabilized. This
new phase is insulating with a gap of 2.7 eV [Fig.
\ref{fig:2}(c)]. Although the carbon atoms in this structure
are threefold coordinated, the structure is stabilized by the
presence of near tetrahedral \emph{sp}$^3$-hybridized
electronic states. For the \emph{sp}$^3$-like hybrids in this
case, there are three covalent $\sigma$-character bonds
connecting each carbon atom to its three nearest neighbors; the
remaining two electrons form a lone-pair orbital, which does
not participate in the direct bonding with other atoms. The
occupied states consist of two distinct groups very similar to
those of cg-N \cite{Chen}. The lowest energy states (more than
-6 eV below the top of the valence band) corresponds to the
$\sigma$ C-C bonding states. The second lowest group, ranging
from -6 eV to the top of valence band, consists of nonbonding
states, which relate to the lone-pair orbital. Although the C-C
bond angle (110$^\circ$) is very close to that in diamond
(109$^\circ$) the bonding scheme is very different, as
discussed in Ref. \cite{Chen}.

In summary, our extensive computational study demonstrates that metallic
linear chains of carbon can be formed by compressing lithium carbide
at relatively low pressure. The conduting properties of these
polyynes-like structures embedded in lithium cages originates from the 
formation of partially filled lone-pair orbitals which favours ionic C-Li bonds
and suppresses the tendency towards a Peierls-like insulating behaviour.
One important observation of the present study is that the chain-like structure
is stable in a pressure range--5-215 GPa--accessible to standardly used high-pressure
equipments, therefore we believe that our findings will encourage immediate
experimental investigations.\\

{\bf Acknowledgement}
The authors thanks R. Podloucky for his
many helpful suggestions and critical reading. Research at Oak
Ridge National Laboratory was sponsored by the Division of
Materials Sciences and Engineering, U. S. Department of Energy
under contract with UT-Battelle, LLC. This research used
resources of the National Energy Research Computing Center,
which is supported by the Office of Science of the US
Department of Energy. Research at the University of Vienna was
supported within the University Focus Research Area {\em
Materials Science} (project ``Multi-scale Simulations of
Materials Properties and Processes in Materials'').\\

\end{document}